\documentstyle{mn}

   \title[bubbles, star clusters and outflow in NGC 3256]
         {The nearest extreme starburst:  bubbles, young star
         clusters, and outflow, in NGC 3256
}

\author[L\'{\i}pari et al.]
   {S.L. L\'{\i}pari$^{1}$, R.J. D\'{\i}az$^{1}$, J.C. Forte$^{2}$,
    R. Terlevich$^{3}$, Y. Taniguchi$^{4}$, M.P. Aguero$^{1}$,      
    \newauthor
    A. Alonso-Herrero$^{5}$, E. Mediavilla$^{6}$, S. Zepf$^{7}$\\
$^{1}$ C\'ordoba Observatory and CONICET, Laprida 854, 5000 C\'ordoba, Argentina.\\
$^{2}$ La Plata Observatory and CONICET, La Plata, Argentina.\\
$^{3}$ Institute of Astronomy, Madingley Road, Cambridge CB3 OHA.\\
$^{4}$ Astronomical Institute, Tohoku University, Aoba, Sendai 980--8578, Japan.\\
$^{5}$ Steward Observatory, University of Arizona, Tucson, AZ 85721, USA.\\
$^{6}$ Instituto de Astrofisica de Canarias, 38205 La Laguna, Tenerife, Spain.\\
$^{7}$ Depart. of Physics and Astronomy, Michigan State Univ.,East Lansing,  MI 48824, USA.
}

\date{Received     ;
      in original form }

\pagerange{\pageref{firstpage}--\pageref{lastpage}}
\pubyear{2002}

\begin{document}

\maketitle

\label{firstpage}

\begin{abstract}

In this work we report, for the extreme starburst in the IR merger
NGC\,3256:

(i) The detection  of 4 galactic bubbles (using {\itshape ESO\/}
NTT and {\itshape HST\/} WFPC2 H$\alpha$--images).
These shells would be associated with previous SNe explosions.

(ii) The first analysis of the spatial distribution of young star clusters
(YSCs) candidates shows that
more than 90 per cent of them are located
in a complex starburst structure, including some of the bubble walls,
three nuclei and three blue asymmetrical spiral arms.

(iii) The kinematics study of the ionized gas in the core of the main
optical nucleus, performed with {\itshape HST\/} STIS spectra.
The shape of the rotation curve
and the emission line profile could be explained  by the presence in the
core of YSCs with out--flow. Any low luminosity AGN associated to this
core would have a mass less than $10^7$\,M$_{\odot}$.
It is also probable that the compact X-ray and radio emission
of ULX(7)N--detected in the main optical nucleus core--is associated to a few recent
SN remnants.

\end{abstract}


\section{INTRODUCTION}\label{introduction}

The study of {\it extreme star formation and the associated galactic
winds processes} is an important issue in modern astrophysics.
In particular, this issue plays an important role in the formation
and evolution of galaxies, at high and low redshift
(see for references Taniguchi \& Shioya 2000; Lipari et al. 2004a,b,c).
Motivated by this, we began a project of study of nearby star
forming + GW galaxies and distant Ly$\alpha$ emitters (for references see 
Lipari et al. 2004a,b,c).
The first step for this project is to understand the extreme star formation
process in nearby galaxies because we can obtain more detailed and
unambiguous information.
Thus, our group started a  study of nearby IR mergers/QSOs,
with galactic winds.

Luminous infrared (IR) galaxies (L$_{IR[8-1000 \mu m]} \geq 10^{11}
L_{\odot}$,
LIRGs) are dusty strong IR emitters -L$_{IR}$/L$_{B} \sim$ 5--300-
where frequently a strong enhancement of star formation is taking place,
and a very high per cent of LIRGs (70--95$\%$) are mergers.
Therefore, at low z, luminous IR mergers (z $\leq$ 0.1) and IR QSOs
(z $\leq$ 0.5) are excellent laboratories for the study of extreme
star formation processes. In these systems very high star formation
and
supernova (SN) rates are expected.
Hubble Space Telescope ({\itshape HST\/}) observations
confirmed that star cluster forming frequency
is highest during violent burst of star formation, in IR
mergers (for references see the reviews of Schweizer 2002;
and Whitmore 2001). Recently, Lipari et al. (2003) reported several galactic
shells and arcs, in nearby IR mergers and IR QSOs with galactic winds.
SNe, shells/arcs and young star clusters are thus good
tracers of extreme starburst events.

NGC~3256 is a nearby LIRG (z = 0.0094) which is a  merger
between--at least--two gas rich disk galaxies
(Toomre 1977; Joseph \& Wright 1985).
The total IR luminosity of NGC~3256 is L$_{IR[8-1000 \mu m]} = 3.3 \times
10^{11} L_{\odot}$, consequently the bolometric luminosity 
makes this system the most luminous nearby galaxy (for V $\leq$ 3000 
km~s$^{-1}$; Sargent, Sanders, \& Phillips 1989). In particular, the
strong 10$\mu$m IR emission in NGC~3256 is {\it ``very extended"}, with 
most of it originating outside the central kpc, indicating  clearly that 
the near to mid--IR emission is probably powered by a recent {\it starburst and
not by a compact object or an AGN} (Graham et al. 1984; Joseph \& Wright 1985).
Throughout the paper, a Hubble constant of H$_{0}$ = 75 km~s$^{-1}$
Mpc$^{-1}$ will be assumed. 
For this merger, we adopted the distance of 37.5 Mpc (V$_{Syst}$ = 2817 
$\pm$15 km~s$^{-1}$; Lipari et al. 2000), and thus, the angular scale is 1$'' 
\approx$180 pc.

\section{OBSERVATIONS}
\label{obser}

Optical {\itshape HST\/} Wide Field Planetary Camera 2 (WFPC2) observations
were analysed, which include broad-band images positioned on the
Planetary Camera (PC) chip with scale of 0\farcs046\,pixel$^{-1}$,
using the filters
F450W (4521 \AA, $\Delta\lambda$ 958 \AA, $\sim$B Cousin filter),
F555W (5407 \AA, $\Delta\lambda$ 1223 \AA, $\sim$V), and
F814W (7940 \AA, $\Delta\lambda$ 1758 \AA, $\sim$I).
In addition, narrow band line ramp filter  LRF--FR680N  images
(6631.4 \AA, $\Delta\lambda$ 76 \AA) were obtained,
positioned in the WFPC2--WF2 CCD. They allowed to map H$\alpha$+[N{\sc ii}]
line emission, in  an effective field of the ramp filter of $\sim$10$''
\times$ 10$''$, with a scale of 0\farcs1\,pixel$^{-1}$.

{\itshape HST\/} Near IR Camera and Multi Object
Spectrometer (NICMOS) archival data
were studied, which include mainly images with the filters 
F160W (1.60 $\mu$m, $\Delta\lambda$ 0.4 $\mu$m), 
F190N (1.90 $\mu$m, $\Delta\lambda$ 0.01 $\mu$m, Pa\,$\alpha$),
F222M (2.20 $\mu$m, $\Delta\lambda$ 0.14 $\mu$m) and
F237M (2.37 $\mu$m, $\Delta\lambda$ 0.14 $\mu$m); using the camera 2
with a scale of 0\farcs076\,pixel$^{-1}$.

Near--UV and optical {\itshape HST\/} Advanced Camera for Surveys (ACS)
archival images were analysed, obtained with the High Resolution Channel
(HRC). They include images with the filters
F220W (2228 \AA, $\Delta\lambda$ 485 \AA),
F330W (3354 \AA, $\Delta\lambda$ 588 \AA),
F555W (5346 \AA, $\Delta\lambda$ 1193 \AA), and
F814W (8333 \AA, $\Delta\lambda$ 2511 \AA).
The scale is 0\farcs027\,pixel$^{-1}$.

{\itshape HST\/} STIS long slit spectroscopy of the main optical nucleus 
of NGC 3256 was obtained. The spectra were taken with the
G750M grating, with the slit  0\farcs1 $\times$ 52\farcs0,
giving a resolution of 50 km s$^{-1}$.
The STIS CCD  has a scale of 0\farcs05 pixel$^{-1}$.
The slit was aligned at PA 91$^{\circ}$.

The {\itshape ESO\/} NTT--SUSI data  include H$\alpha$+[N {\sc ii}]
narrow--band images, using the {\itshape ESO\/} filters
E629 (6571 \AA, $\Delta\lambda$ 115 \AA, H$\alpha$--redshifted) and
E443 (7018 \AA, $\Delta\lambda$ 65 \AA, H$\alpha$--continuum).
These NTT images have seeing of 0\farcs7  FWHM; and the CCD  scale
is 0\farcs13  pixel$^{-1}$. 

The IRAF and STSDAS\footnote{IRAF and STSDAS are the reduction
and analysis software facility developed by NOAO and STScI, respectively}
software packages were used to analyse the {\itshape HST\/} and
{\itshape ESO\/} imaging and spectrophotometric data.
The {\itshape HST\/} data were reduced at the Space Telescope Science
Institute (STScI), using the standard process.
All the bands were calibrated according to the procedures detailed in
Lipari et al. (2004a). Specifically, K--band photometric calibration of
the {\itshape HST\/} NICMOS was performed using
published photometric solution for the F222M and F237M filters
({\itshape HST\/} NICMOS Data Handbook, version 4.0).

\section{RESULTS AND DISCUSSION} \label{resultsdis}

\subsection{Galactic bubbles or shells and the morphology of the extreme star
formation process}
\label{results-sb}

Using {\itshape HST\/} WFPC2 and {\itshape ESO\/} NTT data,
L\'{\i}pari et al. (2000) found that the extreme massive star
formation process shows extended triple asymmetrical spiral arms
(r $\sim$ 5 kpc), emanating from three different nuclei.
We already proposed, that probably the interaction between dynamical
effects of the merger process, the galactic--wind (outflow), and the
molecular+ionized gas (in the inflow phase) could be
the mechanism that generates the  ``massive star formation process
(at scales of 5-6 kpc) with this triple asymmetrical spiral arms
structure".

Figures 1a and 1b show {\itshape ESO\/} NTT  and {\itshape HST\/}
WFPC2--WF2 H$\alpha$--images of this starburst structure.
The main associations of giant H{\sc ii} regions
were labelled according to our previous work.
Figures 1a and 1b depict at least 4 interesting giant galactic shells
or bubbles:

\begin{enumerate}
\item
{\it Shell A:}
This is the more defined shell detected in NGC 3256,
it is located to the south--west of the region 2 (R2: a possible
second optical nucleus; Lipari et al. 2000).
This shell is clearly observed in the NTT H$\alpha$ image with a resolution
of 0\farcs7 (Fig. 1a), however more interesting details 
were detected in the {\itshape HST\/} WFPC2 and ACS images (Fig. 1b).
Half of the circular shell was detected, and the another half is
heavily obscured by a strong and wide dust lines system.
The near IR NICMOS data confirm this last proposition, the bubble wall
is more extended to the south at this bands, and shows a knotty structure.
This obscured area is located in the south region of the main body of this
merger (Lipari et al. 2000).
The diameter of this galactic shell  is $D_A\,=1\farcs8=324$\,pc.

The  {\itshape HST\/} ACS  F220W and F330W UV--images 
show remarkable details of the blue compact knots located
at the border of this shell (which are brighter in the east side).
This knots depict strong UV continuum, probably associated to the
presence of a high number of massive O and B stars.

{\itshape HST\/} WFPC2--H$\alpha$, F450W, ACS--F330W, F555W, F814W and
NICMOS--F222M, F237M, Pa$\alpha$ images show exactly at the centre of
this shell,
a young star cluster (YSC) candidate plus a giant H{\sc ii} region.
This configuration constraints the age of the starforming complex to
a few Myr, this kind of coincidences between H{\sc ii} region
emissions and YSCs (for the most massive clusters of
$\sim$10$^{6-7}$ M$_{\odot}$) could only be detected during the first
7 $\times$ 10$^6$ yr (Alonso-Herrero et al. 2002).

\item
{\it Bubble B:}
Two narrow arcs or shells were found in the west side of the region 2
(each one is marked with a line in Fig. 1b).
Furthermore, {\itshape HST\/} ACS F220W and F330W UV--images show that
the core of this region (R2) is the brightest area of NGC 3256, in the near--UV.
This fact suggests the presence of a very high number of massive O and B hot
stars, which normally start the first phase of the galactic wind/bubble
and also are the progenitors of core--collapse SNe.

The bubble structure detected around R2, was already noted by Lira et al.
(2002).
The diameter of the more external arc is about the same than Shell A: 
$D_B\,=1\farcs9=342$\,pc.

\item
{\it Shell C:}
A less extended arc was detected to the west side of region 4.
It is observed  more clearly
in {\itshape HST\/} WF2 H$\alpha$ data (Fig. 1b).
The diameter of this shell is $D_C\,=0\farcs9=162$\,pc.

\item
{\it Shell D:}
To the west, and adjacent to region 7, the NTT H$\alpha$ image (Fig. 1a)
shows another structure with bubble shape.
However, more clear details of this structure are observed in the
{\itshape HST\/} WFPC2 and ACS images, showing  several blue compact knots.
These knots are located at the inner border of the shell. 
The diameter of this bubble is $D_D\,=1\farcs7=305$\,pc.

\end{enumerate}

These  ``shells, bubbles, arcs" could be associated mainly to
explosions of SNe and the final phase of the galactic--wind, i.e., the
blow--out phase of  galactic bubbles (Norman \& Ikeuchi 1989).
Multiple explosion of SNe, from massive progenitors, are the main
galactic objects capable to generate this blow--out phase
(Norman \& Ikeuchi 1989). Massive YSCs and associations of giant H{\sc ii}
regions are the  places, where multiple SN explosions are expected.

\subsection{The young star clusters in NGC 3256 ({\itshape HST\/} WFPC2,
NICMOS and ACS images)}
\label{results-ysc}

We have started a multi--wavelength studies of the young stellar
cluster candidates (YSCc) in NGC 3256, using {\itshape HST\/} ACS/UV,
WFPC2/Optical, NICMOS/near-IR images. From these {\itshape HST\/} data we  
study the colours, ages, masses, subgroups, spatial distribution and
metallicities of these YSCc. Here we report new results from this study.

First, the colour distribution, the colour--magnitude diagram and the
spatial distribution of YSCc (of NGC 3256) were analysed.
Using mainly the Catalogue of YSCc published by Zepf et al. (1999: their
Table 1). We note that
from this catalogue, the positions of YSCc in pixels (and also the
corrected values of RA and DEC) were used; since the published data of
RA and DEC have small errors, which were generated in the transformation
from pixels to RA--DEC. An electronic corrected version of this Catalogue of
YSCc is available at zepf@pa.msu.edu and lipari@oac.uncor.edu.

From this study, the following results were found:

\begin{enumerate}

\item
{\it Complete sample of YSCc:} The optical spatial distribution of all
the YSCc was first analysed.
Fig. 2a depicts their interesting distribution. We found that
the YSCc are located in more than a 90 per cent in the starburst
regions: they concentrate in the 3 arms, 3 nuclei, and in some of the
bubble walls, tracing the extreme starburst.

For the southern obscured areas, especially for the
obscured southern nucleus (R3),
and for the obscured part of the arm III, we found a similar
result. In particular, using our near--IR Catalogue of YSCc and H{\sc ii}
regions (with objects detected in the HST NICMOS H and Pa$\alpha$ bands
images; Alonso--Herrero et al. 2002), we also found that the YSCc and
H{\sc ii} regions are located mainly in the nuclei, the blue arms
and, at some of the bubble walls.

\item
{\it Very Red YSCc:} The colour (B--I) distribution of the YSC candidates
shows bimodal shape with a main (strong: $\sim$920 objects) peak at
(B--I) = 0.75 and a second (small: $\sim$40 objects) peak at (B--I) = 2.6.
Thus, a first subgroup of YSCc was selected
including the very red objects found around this second peak.
These YSCc were selected using the condition (B--I) $>$ 2.3,
since at (B--I) = 2.3 starts the distribution of YSCc associated with
this second peak.
Fig. 2b shows that these very red YSCc are located mainly in the obscured
arm III and around the core of the main nucleus (at a radius r $\sim$0\farcs8
= 144 pc).

Zhang, Fall, \& Withmore (2001) found a similar subgroup of red YSCc, in
the IR pre-merger ``The Antennae".
They suggested that these objects could be the youngest clusters, detected
when they emerge from their dust cocoons.
For NGC 3256, the location of the very red YSCc suggest that these clusters
candidate are  mainly very dusty objects.

\item
{\it Blue and Very Blue YSCc:} The observed and theoretical YSCc
colour--magnitude diagrams (B--I) vs. m$_B$ (Zepf et al. 1999: their Figs.
3 and 4), clearly show that
the stellar population models--of Bruzual \& Charlot (2003) and
Leitherer et al. (1999)--generate unique ages between 3 and 7 Myr, for
YSCc with colour in the range --0.55 $<$ (B--I) $<$ 0.5. See for details,
English \& Freeman (2003).

Using this fact, we selected two subgroups of blue and very blue YSCc: those
with (B--I) $<$ 0.5 and (B--I) $<$ 0.0, respectively.
Thus the first subgroup include the second one.
Fig. 2c  depicts the spatial distribution of these two subgroups.
Both samples were found distributed mainly in the second
optical nucleus (R2) and the closer asymmetrical arm II. This
fact is more evident for the very blue YSCc (Fig. 2c).

\end{enumerate}

These results are in  agreement with a ``composite" multiple merger
model proposed for NGC 3256 (Lipari et al. 2000).
In this model the main merger could be between two gas rich massive
spiral galaxies (Sc), and then with a satellite galaxy (of one of these Sc
systems). In particular,
we proposed that the kinematics and spectroscopic properties of the core
of region 2 could be associated to a 3rd. nucleus, arisen in a satellite
of the two original spiral galaxies that collided.
In order to explore these type of scenarios we are working in detailed
numerical simulations, which are in progress (Lipari et al. 2000, 2004, in
preparation). The  results of these kind of simulations
suggest that the two massive galaxies are the main component of the
merger event, and that the satellite generally merges later.
Since the starbursts in each blue asymmetrical arms are probably related with
the multiple merger event and associated with each nuclei, we already
suggested that
the starburst could be younger in R2 and the arm II (Lipari et al. 2000).
The new results presented in this work support this suggestion: (i) the very
blue YSCc are located mainly in the region 2 and the arm II; and (ii) the
brightest near--UV areas (in the {\itshape HST\/} ACS images) are also
located in the region 2 and the arm II.

It should be noted that $\sim$5\% of the YSCc studied are objects with
absolute magnitude less than -13 (luminosities greater than
10$^{40}$\,erg\,s$^{-1}$) which are strong candidates to be massive YSC
with masses in the range $\sim10^{6-7}$\,M$_{\odot}$.

It is interesting to remark that a simple estimation of the kinematical energy
of the Shells A and B, using a wall thickness of 20-40 pc, electron density
N$_e\sim1000$\,cm$^{-3}$ (Lipari et al. 2000), and filling factor $\sim0.1$,
yields 3-6$\times10^{52}$\,erg.  This is equivalent to 1/3 of the kinetic
energy of the nuclear outflow in M\,82 (Bland \& Tully 1988).

\subsection{The core kinematics and the outflow of the main optical
nucleus ({\itshape HST\/} STIS spectra)}        
\label{results-r1}

HST--STIS  $0\farcs1$ $\times$ $52\farcs0$ long slit spectra show a
rotational pattern in the central arcsecond (core) of NGC\,3256 main
optical nucleus (Fig. 3a), with an amplitude of
$\Delta\,V_r=160$\,km\,s$^{-1}$. For this core, Neff et al. (2003)
suggested that a compact X-ray and radio source
ULX(7) North (Lira et al. 2002; our R1 region) would have a mass of
at least 10$^8$\,M$_{\odot}$ and could be associated to a supermassive
black hole.
Nevertheless, another possibility must be considered, since there is
an extranuclear compact X-ray and radio source, ULX(13) [which is
located inside of the the H{\sc ii} region 5, called R5 in our
Fig. 1a],
with the same spectral properties than ULX(7)N but not associated to
a nuclear mass concentration.

In addition, using {\itshape HST\/} STIS long slit spectra with the slit
$0\farcs2$ $\times$ $52\farcs0$  (100 km s$^{-1}$ of resolution),
Neff et al. (2003) suggested
the presence, of  broadened  H$\alpha$ and [N{\sc ii}]
emission lines, with linewidths of $\sim$450 km s$^{-1}$, as evidence for
an AGN in the core. Figure 4  shows the {\itshape HST\/} STIS spectrum of
the core, using the $0\farcs1$ $\times$ $52\farcs0$ long slit (giving a
better spectral resolution: 50 km s$^{-1}$).
From this figure is very clear that the broadening of the
H$\alpha$ and [N{\sc ii}] emission lines is due to the presence of 
a strong blue out--flow component. This component was previously detected
at large scale, and
associated to a nuclear galactic wind, by Lipari et al. (2000).
Moreover, this broad out--flow component was observed
also in the spectra located at several hundred pc from the nucleus.
This fact rules out the possibility that this emission could be associated
with the broad line region of an AGN.  In the core and in several
STIS spectra (with $0\farcs1$ slit) we measure for the
main component of the emission lines H$\alpha$ and [N{\sc ii}],  FWHM in
the range  90--130 km s$^{-1}$, these are normal values for H{\sc ii}
regions. Meanwhile, for the out-flow component was measured  FWHM in the
range  200--260 km s$^{-1}$, probably associated with shock processes.

The central rotation curve of region 1 (Figure 3a) is well fitted by a
Plummer--Kuzmin law (Binney \& Tremaine 1987), that could correspond to a
spherical-- or disk--like mass distribution with the following parameters:
$V_{sys}=2898$\,km\,s$^{-1}$, scale radius $a=40$\,pc,
$M_{tot}=2.16\times10^8$\,M$_{\odot}$.  Moreover, any single component
fitted model cannot have a scale radius smaller than 40\,pc
(the resolution is better than 18 pc).
In fact, the central slope and maximum of the rotation curve could not
be fitted by a point--mass larger than $\sim10^7$\,M$_{\odot}$, even
allowing for a beam smearing corresponding to a resolution $0\farcs15$,
which is twice the slit width and the image quality of the STIS data
used. This is well illustrated by the grey line in Figure 3a, which
corresponds to a point central mass of $2\times10^7$\,M$_{\odot}$,
with a beam smearing of $0\farcs15$.

Considering that several young star cluster candidates of mass
$\sim10^{6-7}$\,M$_{\odot}$
and luminosities of $\sim10^{39-40}$\,erg\,s$^{-1}$ are observed inside
$r\sim50$\,pc and that the rotation curve corresponds to that of an extended
mass distribution, it seems likely that any low luminosity AGN associated to
ULX(7)N would have a mass less than $\sim10^7$\,M$_{\odot}$ and would  not
contribute significatively to the core luminosity.
Indeed, the radial change of mass to F814W($\sim$I)--luminosity ratio
(Figure 3b), calculated in the way described in detail
in L\'ipari et al. (2004a), does not allow the possibility of a hidden dark
mass kinematically detected, down to the resolution of $0\farcs1$.
There is no evidence of a strong central light or gravity
point source, that would generate, respectively, a strong deep or peak
in the M/L$_I$ profile.
Moreover, the central values of the mass to F814W($\sim$I)--luminosity
ratio are similar to those found in other extended starbursts, e.g. our
study of M/L$_I$ in NGC 2623.

It seems more plausible that the compact X-ray and radio emission of ULX(7)N,
as for ULX(13), is associated to a few recent SN remnants in the massive
young star clusters of the nucleus (R1) core.  As was pointed out by
Neff et al. (2003) this would require several supernovae in the nucleus
in the last 15 years, which seems a possibility to be considered.

\subsection{The possible  IR supernova in 1997
{\itshape HST\/} NICMOS images (in NGC 3256)}
\label{results-sn}

Using the  {\itshape HST\/} NICMOS images, taken at F222M ($\sim$K,
2.2--$\mu$m) and F237M (2.4--$\mu$m) bands we found a very bright 
source in the main body of NGC 3256. This object
appears in the {\itshape HST\/} images obtained in 1997 Nov. 28.9 UT.
We previously identified this object as a possible SN event (PSN9711-001;
Lipari et al. 2004d). Recently, we found that this object is associated
to thermal emission from the coronagraphy hole of NIC-2 camera.

\section*{Acknowledgments}

Part of the {\itshape HST\/} observations were obtained from the archives at
{\itshape ESO\/} Garching and STScI Baltimore.
The authors thank  J. Acosta--Pulido, H. Dottori, and B. Garcia--Lorenzo
 for discussions and assistance.
Finally, we wish to thank the referee, Dr. Richard de Grijs, for
constructive and valuable comments, which helped to improve the content
and presentation of the paper.

\clearpage

\begin{figure*}
\vspace{12.0 cm}
\begin{tabular}{c}
\includegraphics{fig1a.ps} \cr
\end{tabular}
\vspace{7.0 cm}
\caption {
{\itshape ESO\/} NTT--SUSI and {\itshape HST\/} WFPC2--WF2 H$\alpha$--images,
of the central region of NGC 3256.
Letter colour Plate 1.
}
\label{bub1}
\end{figure*}

\clearpage

\begin{figure*}
\vspace{12.0 cm}
\begin{tabular}{c}
\includegraphics{fig1b.ps} \cr
\end{tabular}
\vspace{7.0 cm}
\addtocounter{figure}{-1}
\caption {
Continued.
}
\label{bub2}
\end{figure*}

\clearpage

\begin{figure*}
\vspace{12.0 cm}
\begin{tabular}{c}
\includegraphics{fig2a.ps} \cr
\includegraphics{fig2b.ps} \cr
\end{tabular}
\vspace{9.0 cm}
\caption {
{\itshape HST\/} WFPC2--PC F435W ($\sim$B) image of the central region of
NGC 3256, superimposed to:
(a) the total sample of young stellar cluster (YSC) candidates,
(b) the sub-sample of very red YSC candidates, 
(c) the sub-sample of blue and very blue YSC candidates
(with blue and black colour, respectivelly).
For details see the text.
The crosses mark the positions of the YSC candidates.
Letter colour Plate 2.
}
\label{ysc1}
\end{figure*}

\clearpage

\begin{figure*}
\vspace{12.0 cm}
\begin{tabular}{c}
\includegraphics{fig2c.ps} \cr
\end{tabular}
\vspace{9.0 cm}
\addtocounter{figure}{-1}
\caption {
Continued.
}
\label{ysc2}
\end{figure*}

\clearpage

\begin{figure*}
\vspace{12.0 cm}
\begin{tabular}{c}
\includegraphics{fig3a.ps} \cr
\end{tabular}
\vspace{7.0 cm}
\caption {
(a) Velocity values of the ionized gas obtained for the core of the main
optical nucleus. They were obtained from long slit {\itshape HST\/}
STIS spectra, at PA = 91$^{\circ}$, i.e. close to the kinematics major axis.
The solid line shows the best Plummer--Kuzmin potential fitted, and the grey
line corresponds to a point central mass of $2\times10^7$\,M$_{\odot}$,
with a beam smearing of $0\farcs15$;
(b) the mass to {\itshape HST\/} WFPC2 F814W($\sim$I)--luminosity ratio.
}
\label{k1}
\end{figure*}

\clearpage

\begin{figure*}
\vspace{12.0 cm}
\begin{tabular}{c}
\includegraphics{fig3b.ps} \cr
\end{tabular}
\vspace{7.0 cm}
\addtocounter{figure}{-1}
\caption {
Continued.
}
\label{k2}
\end{figure*}

\clearpage

\begin{figure*}
\vspace{12.0 cm}
\begin{tabular}{c}
\includegraphics{fig4.ps} \cr
\end{tabular}
\vspace{7.0 cm}
\caption {
Long slit {\itshape HST\/} STIS spectrum of the main optical
nucleus core of NGC 3256 (at PA = 91$^{\circ}$).
It shows the H$\alpha$+[N{\sc ii}] emission line blend, with a clear
blue out--flow component (in each emission line).
}
\label{k3}
\end{figure*}

\end{document}